\documentclass[letterpaper,journal]{IEEEtran}
\usepackage{amsmath,amsfonts}
\usepackage{array}
\usepackage{booktabs}
\usepackage{multirow}
\usepackage[caption=false,font=footnotesize]{subfig}
\usepackage{stfloats}
\usepackage{url}
\usepackage{graphicx}
\usepackage{cite}
\usepackage{xcolor}
\usepackage{enumitem}
\usepackage{etoolbox}
\usepackage{hyperref}
\hypersetup{hidelinks}
\makeatletter
\patchcmd{\thebibliography}{\footnotesize}{\scriptsize}{}{}
\makeatother

\AtBeginDocument{%
  }

\begin{document}

\title{AT-ADD: A Benchmark and Challenge for Robust and All-Type Audio Deepfake Detection}

\author{Yuankun~Xie,
Haonan~Cheng,
Jiayi~Zhou,
Xiaoxuan~Guo,
Tao~Wang,
Changhao~Zhang, Jian~Liu, Weiqiang~Wang,
Ruibo~Fu,
Xiaopeng~Wang,
Hengyan~Huang, Xiaoying~Huang,
Long~Ye,
Guangtao~Zhai,
and Kong~Aik~Lee%

\thanks{Yuankun Xie, Haonan Cheng, Xiaoxuan Guo, Hengyan Huang, Xiaoying Huang, and Long Ye are with the Communication University of China, Beijing, China. Yuankun Xie and Xiaoxuan Guo are also with Ant Group.}%
\thanks{Jiayi Zhou, Tao Wang, Changhao Zhang, Jian Liu, and Weiqiang Wang are with Machine Intelligence, Ant Group, Shanghai, China.}%
\thanks{Ruibo Fu is with the Institute of Automation, Chinese Academy of Sciences, Beijing, China. Xiaopeng Wang is with Beijing Institute of Technology, Beijing, China. Guangtao Zhai is with Shanghai Jiao Tong University, Shanghai, China.}%
\thanks{Yuankun Xie and Kong Aik Lee are also with The Hong Kong Polytechnic University, Hong Kong SAR, China.}%
\thanks{Official website: \url{https://at-add.com}.}}

\markboth{JOURNAL OF \LaTeX\ CLASS FILES, VOL. 14, NO. 8, AUGUST 2021}%
{Xie \MakeLowercase{\textit{et al.}}: AT-ADD Benchmark and Challenge}
\maketitle

\begin{abstract}
Recent audio generation models can synthesize high-fidelity speech, environmental sound, singing voice, and music, creating new risks for multimedia trust. Existing audio deepfake detection (ADD) benchmarks remain predominantly speech-centric and often under-represent realistic channel variation and heterogeneous audio types. This paper presents all-type ADD (AT-ADD), a large-scale benchmark and challenge designed to evaluate both robust speech deepfake detection and all-type audio deepfake detection. Track~1 evaluates speech deepfake detection under unseen synthesizers, diverse recording conditions, signal perturbations, and replay effects. Track~2 evaluates type-agnostic deepfake detection over speech, sound, singing, and music when the audio type is unknown at test time. We detail the dataset construction, evaluation protocol, and reproducible baselines, and analyze the final systems submitted to the ACM Multimedia 2026 Grand Challenge. The strongest official baseline obtains 76.73\% and 79.47\% Macro-F1 on the Track~1 and Track~2 evaluation sets, respectively, whereas the winning challenge systems reach 90.71\% and 96.10\%. Beyond the aggregate rankings, sample-level analysis of the top five submissions reveals concentrated failures on specific synthesis methods and replayed speech. It also quantifies cross-system error complementarity and separates stable leaderboard gaps from statistically ambiguous adjacent ranks. Our results show that large-scale self-supervised representations, condition-aware augmentation, multi-crop inference, and structured fusion or routing are central to generalization, while generator-specific robustness and balanced detection across heterogeneous audio types remain unresolved.
\end{abstract}






\section{Introduction}

\begin{figure}[!tb]
	\centering
	\subfloat{\includegraphics[width=3.3in]{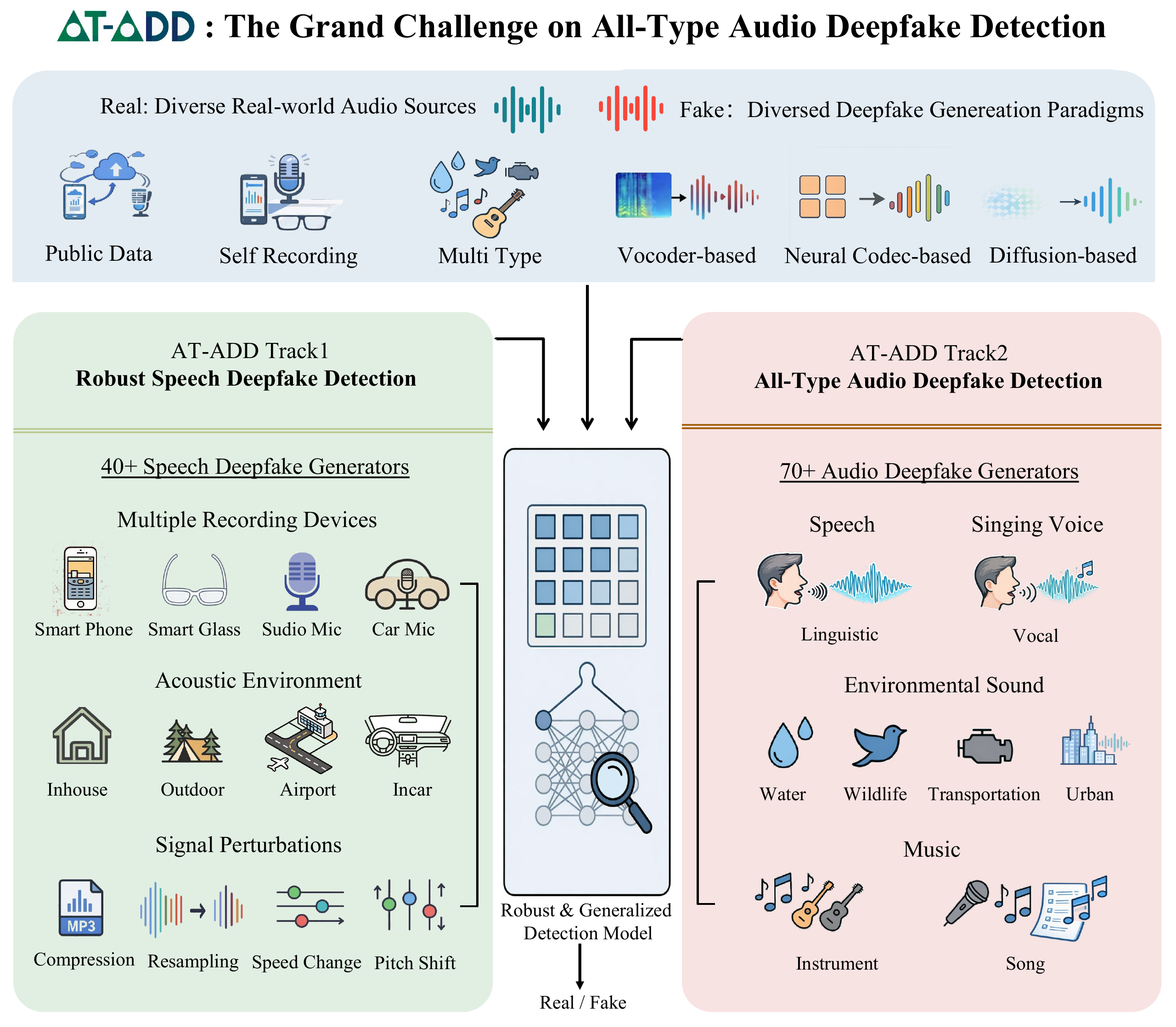}}
	\hfil
	\caption{AT-ADD overview: Track~1 targets robust speech deepfake detection; Track~2 covers speech, sound, singing, and music without test-time audio-type labels.}
	\label{fig:intro} 
\end{figure}

Recent advances in audio generation technologies, particularly Audio Large Language Models (ALLMs)~\cite{borsos2023audiolm}, have significantly improved the realism, scalability, and accessibility of synthetic audio. Modern generative systems are now capable of producing high-fidelity audio across a wide range of content types, including speech~\cite{du2024cosyvoice2}, environmental sounds~\cite{liu2023audioldm}, singing voices~\cite{liu2022diffsinger}, and music~\cite{agostinelli2023musiclm}. While these developments greatly benefit content creation and multimedia applications, they also introduce serious security and trust risks, as audio deepfakes can be generated and disseminated at scale with increasing realism~\cite{wang2024asvspoof}.

Despite growing research efforts in audio deepfake detection (ADD)~\cite{nautsch2021asvspoof,yi2023add}, existing methods and benchmarks remain largely focused on speech and are typically evaluated under relatively controlled conditions. Consequently, current countermeasures (CMs) often rely on speech-specific artifacts and exhibit limited robustness when deployed in real-world scenarios involving channel variability, environmental noise, compression, replay attacks, and other distortions~\cite{liu2023asvspoof,kawa2023defense}. Furthermore, their generalization capability remains insufficient when faced with emerging generation paradigms, such as ALLM-based synthesis and neural codec-driven generation, as well as diverse non-speech audio types~\cite{xie2025codecfake,xie2025detect}.

To address these challenges, we introduce the AT-ADD (All-Type Audio Deepfake Detection) Grand Challenge at ACM Multimedia 2026~\cite{xie2026ataddsummary}. The goal of AT-ADD is to bridge the gap between idealized research settings and real-world multimedia forensics by systematically evaluating both robustness under realistic conditions and generalization across audio types and unseen generation methods.

The challenge comprises two tracks, described as follows.

\textbf{Track 1: Robust Speech Deepfake Detection.}
This track focuses on robustness in real-world speech deepfake detection. We construct a large-scale dataset (AT-ADD Track 1) covering 47 state-of-the-art speech generation models, spanning vocoder-based~\cite{lee2022bigvgan}, neural codec-based~\cite{du2024cosyvoice2}, and diffusion-based~\cite{liu2022diffgantts} paradigms, with particular emphasis on emerging ALLM-driven synthesis. The real speech data cover 11,299 unique speakers drawn from multiple public datasets~\cite{koizumi2023libritts,ardila2020common} and in-the-wild recordings, spanning diverse languages, recording devices (e.g., smartphones, in-vehicle systems, wearable devices), and acoustic environments. To simulate realistic deployment conditions, we further introduce a wide range of degradations, including background noise, reverberation, replay attack, compression, resampling, and speed perturbation. This track evaluates robustness and cross-domain generalization under complex real-world conditions, encouraging models to move beyond narrow artifact cues toward more fundamental and transferable forensic representations.

\textbf{Track 2: All-Type Audio Deepfake Detection.}
Building upon Track 1, this track extends the detection task from speech to all types of audio, including environmental sounds, singing voices, and music. We construct a new dataset (AT-ADD Track 2) covering 68 audio generation models across different generation mechanisms. Unlike Track 1, no additional  signal-level perturbations are introduced in this track, allowing for a more controlled investigation of cross-type generalization. Real audio is collected from multiple public datasets~\cite{mesaros2016tut,wang2022opencpop,agostinelli2023musiclm}, while synthetic audio is generated under a unified framework without introducing additional distortions. Participants are required to design type-agnostic CMs capable of distinguishing real and fake audio under unseen audio categories. This track encourages models to capture shared synthesis artifacts across different audio types, promoting the development of universal audio deepfake detection methods.

Together, the tracks form a progressive framework: Track~1 emphasizes robust speech deepfake detection under realistic conditions, while Track~2 targets generalization across audio types and recording domains. This design addresses evolving threats from speech manipulation to open-domain audio generation.

The main contributions of this work are as follows:
\begin{itemize}[leftmargin=*,nosep]
    \item We construct and publicly release the AT-ADD Track~1\footnote{\url{https://huggingface.co/datasets/xieyuankun/AT-ADD-Track1}} and Track~2\footnote{\url{https://huggingface.co/datasets/xieyuankun/AT-ADD-Track2}} datasets, together with their complete construction protocols and generator-level compositions. 
    \item We establish reproducible baselines\footnote{\url{https://github.com/xieyuankun/AT-ADD-Baseline}} spanning conventional countermeasures, self-supervised learning models, and audio large language models under a unified AT-ADD closed-setting protocol, reporting both aggregated and type-wise performance.
    \item We analyze the performance of the top systems and evaluation-set difficulty, revealing effective system designs and challenging generators and audio types.
\end{itemize}

\section{Challenge Tasks}
\label{sec:tasks}

The AT-ADD challenge is organized into two complementary and progressively structured tracks under a \textbf{closed setting} (fixed condition). This design aims to assess which types of CMs can achieve stronger generalization and robustness under limited data conditions. Participants are required to train their CMs strictly using only the data provided by the organizers, thereby ensuring a fair and controlled comparison across methods.

\subsection{Track 1: Robust Speech Deepfake Detection}
\label{sec:track1}

\textbf{Goal.}
Track~1 aims to bridge the gap between existing benchmarks and real-world deployment scenarios for speech deepfake detection. It evaluates whether a detector can remain reliable under realistic domain shifts and practical post-processing effects, while maintaining strong performance against modern high-fidelity synthesis systems.

\textbf{Task definition.}
Given an input speech utterance, participants are required to predict whether the input is \emph{real} or \emph{fake}. In this task, \emph{fake} refers specifically to deepfake speech generated using deep neural network-based methods, while \emph{real} refers to non-deepfake speech. It should be noted that signal distortions or transformations, such as compression, resampling, speed perturbation, and pitch shifting, as well as replay-based attacks, do not change the original real/fake label in this task. The training and development data are fully provided by the organizers, and the use of external data is not allowed under the closed setting.

The evaluation set includes deepfake samples generated by methods that are \emph{unseen} during training and reflect recent state-of-the-art generation techniques. Meanwhile, the real speech in the evaluation set is collected under realistic conditions, involving variations in recording devices, acoustic environments, languages, and other real-world factors.

\subsection{Track 2: All-Type Audio Deepfake Detection}
\label{sec:track2}

\textbf{Goal.}
Track~2 targets universal audio deepfake detection across heterogeneous audio types and aims to develop \emph{type-agnostic} detectors that generalize across both audio types and unseen generation methods.

\textbf{Task definition.}
Given an input audio clip of unknown type, participants are required to determine whether it is \emph{real} or \emph{fake}. In this task, \emph{fake} denotes deepfake audio generated by deep neural network-based methods, whereas \emph{real} denotes non-deepfake audio. Notably, in Track~2, audio-type labels (i.e., speech, sound, singing, and music) are \emph{not} available at test time, reflecting realistic deployment scenarios.

Similar to Track~1, this track follows a closed setting, where participants must use only the provided training and development data, without access to external resources.

\section{Related Work}

ADD has progressed rapidly with the rise of high-fidelity neural generative models. Existing studies are dominated by speech-focused benchmarks and methods~\cite{nautsch2021asvspoof,yi2023add}, where SSL representations and attention-based back-ends have driven substantial performance gains~\cite{jung2022aasist,tak2022automatic}. In contrast, detection for non-speech audio (sound, singing voice, and music) remains less explored and is still largely benchmark-driven, while generalization across heterogeneous audio types is an emerging yet under-established research frontier. In the following, we review prior work by audio type---speech, sound, singing voice, and music---and then discuss cross-type ADD, which motivates the need for a unified and realistic all-type benchmark.

\begin{table*}[t]
\centering
\caption{Comparison with representative audio deepfake detection challenge datasets.}
\label{tab:challenge_comparison}
\footnotesize
\setlength{\tabcolsep}{6pt}
\renewcommand{\arraystretch}{1.05}
\begin{tabular}{lccccccc}
\toprule
\textbf{Dataset / track} & \textbf{Year} & \textbf{Venue} & \textbf{Speakers} & \textbf{Language} & \textbf{Generators} & \textbf{Types} & \textbf{Clips} \\
\midrule
ASVspoof 2019 LA~\cite{nautsch2021asvspoof} & 2019 & INTERSPEECH & 107 & 1 & 19 & 1 & 121,461 \\
ADD 2022 LF~\cite{yi2022add} & 2022 & ICASSP & 80+ & 1 & -- & 1 & 165,607 \\
ADD 2023 FG-D~\cite{yi2023add} & 2023 & IJCAI & -- & 1 & 22 & 1 & 175,601 \\
ASVspoof 5 T1~\cite{wang2024asvspoof} & 2024 & INTERSPEECH & 1,922 & 1 & 32 & 1 & 1,004,081 \\
SVDD 2024 Ctr~\cite{zhang2024svdd} & 2024 & SLT & 164 & 2 & 14 & 1 & 220,798 \\
ESDD T1~\cite{yin25_interspeech} & 2026 & ICASSP & -- &  -- & 7 & 1 & 184,765 \\
ESDD 2~\cite{zhang2026esdd2} & 2026 & ICME & -- & -- & 6+ & 2 & 255,433 \\
RADAR 2026~\cite{luong2026radar} & 2026 & APSIPA & -- & 4 & 15 & 1 & 146,760 \\
\textbf{AT-ADD T1} & \textbf{2026} & \textbf{ACM MM} & \textbf{11,299} & \textbf{96} & \textbf{47} & \textbf{1} & \textbf{245,655} \\
\textbf{AT-ADD T2} & \textbf{2026} & \textbf{ACM MM} & \textbf{11,299+} & \textbf{96+} & \textbf{68} & \textbf{4} & \textbf{467,223} \\
\bottomrule
\end{tabular}
\vspace{1pt}

\parbox{0.98\textwidth}{\scriptsize ``--'' denotes a statistic not reported in the official protocol, whereas N/A denotes a field that is not applicable. Statistics refer to the main detection track and its final evaluation round; progress subsets sampled from evaluation data are counted only once. Speaker counts refer to identified vocal identities. Language counts use base language codes, grouping regional variants; Track~2 inherits at least the 96 languages in its speech subset. A plus sign denotes a reported lower bound.}
\end{table*}

\textbf{Speech}. Speech deepfake detection has been extensively studied, largely driven by the ASVspoof challenges \cite{nautsch2021asvspoof, liu2023asvspoof, wang2024asvspoof} and ADD challenges \cite{yi2022add, yi2023add}. Representative CMs include AASIST \cite{jung2022aasist} and SSL-based pipelines that combine XLSR with AASIST \cite{tak2022automatic}. Subsequent studies have investigated different SSL representations \cite{phukan2024heterogeneity,kheir2025comprehensive}, layer utilization of SSL features \cite{zhang2024audio,wang2025mixture,pan2024attentive}, and robustness \cite{zhang2025i,xu2025alden,kawa2023defense}. However, a substantial gap remains between existing public benchmarks and real-world conditions (e.g., diverse capture devices, channel effects, and replay attack \cite{muller2025replay, zhang2025echofake}), highlighting the need for new datasets and protocols that better reflect realistic deployment scenarios and enable more faithful evaluation of detection performance in the wild.

\textbf{Sound}. Compared to speech deepfake detection, research on  environmental sound deepfake detection is still relatively nascent and is largely driven by dataset and benchmark construction. The Environmental Sound Deepfake Detection (ESDD) Challenge \cite{yin25_interspeech} has recently advanced this area by covering a wide range of ALLM-based text-to-audio (TTA) and audio-to-audio (ATA) synthesis methods. Current state-of-the-art solutions typically leverage sound-oriented SSL representations such as SSLAM \cite{guo2025envsslam}.

\textbf{Singing voice}. Singing voice can be considered a subcategory of music; however, it is treated as a distinct audio type in this challenge due to its unique characteristics and the high difficulty of deepfake song detection. Unlike general music, singing voice shares strong similarities with speech as both are produced by human vocal mechanisms, while also exhibiting complex musical structures such as melody and rhythm. These properties make singing voice particularly challenging for existing CMs. Recent work, such as SVDD \cite{zhang2024svdd}, has promoted research in deepfake singing voice detection. Competitive approaches often leverage hybrid representations by combining speech-oriented SSL features (e.g., XLSR) with music-oriented SSL models such as MERT \cite{li2024mert} and WavLM \cite{chen2022wavlm}, as explored in recent studies \cite{10832226,zhang2024xwsb,chen2024singing}.

\textbf{Music}. Music represents a broad and diverse audio type that encompasses both instrumental compositions and songs. Compared to speech and singing voice, music exhibits higher variability in structure, timbre, and generation mechanisms, posing additional challenges for deepfake detection.
FakeMusicCaps \cite{comanducci2024fakemusiccaps} provides a benchmark for synthetic-music detection and enables the study of text-to-music (TTM) generation artifacts. However, methodological explorations remain relatively limited compared to speech \cite{li2024detecting,wei2025voices}.

\textbf{Cross-type.} A few studies have investigated transfer across audio types, for example between speech and singing voice \cite{gohari2025audio}, from speech to music \cite{li2024audio}, and ESDD 2 (from sound to both speech and sound deepfake detection settings \cite{zhang2026esdd2}). Xie et al.~\cite{xie2025detect} further establish an SSL-based benchmark for all-type ADD and propose wavelet prompt tuning to improve cross-type generalization. However, these studies are still grounded on relatively limited and task-specific datasets, and the community is still lacking a comprehensive and widely accepted benchmark that systematically covers the full spectrum of audio types and realistic conditions.

Overall, despite substantial progress in speech deepfake detection, a clear gap remains between academic benchmarks and real-world deployment, particularly in terms of robustness to complex acoustic environments, and rapidly evolving deepfake paradigms. This motivates the need for a robust speech deepfake detection benchmark that more faithfully reflects practical scenarios. Meanwhile, research on all-type audio deepfake detection is still at an early stage, and a unified evaluation model spanning speech, sound, singing voice, and music is essential to drive the next generation of generalizable CMs.

Table~\ref{tab:challenge_comparison} compares representative challenge datasets. AT-ADD combines substantially broader speaker coverage with four audio types and diverse generation conditions, complementing the larger but more narrowly scoped prior benchmarks.

\section{Datasets and Resources}

To support the AT-ADD challenge, we construct two benchmark datasets: \textbf{AT-ADD Track~1} and \textbf{AT-ADD Track~2}, corresponding to robust speech deepfake detection and all-type audio deepfake detection, respectively. The datasets and metadata are released on Hugging Face for Track~1 and Track~2. For both tracks, we provide standardized train, development (dev), and evaluation (eval) splits under a closed setting. In addition, a progress subset is provided for progress evaluation, which is sampled from the evaluation set with the same distribution and constitutes 20\% of the full eval set. An overview of the two tracks is presented in Table~\ref{tab:alldata_protocol}.

Detailed dataset compositions and statistics are presented in the following subsections. It should be noted that, to ensure data quality (e.g., by removing fully silent segments), we applied a series of screening and filtering procedures. As a result, the number of samples in each condition is not perfectly uniform. 
\begin{table*}[t]
\centering
\caption{AT-ADD statistics (number of clips) for Track 1 and Track 2.}
\label{tab:alldata_protocol}
\begin{tabular}{c c | ccccc}
\hline
\multirow{2}{*}{\textbf{Split}} & 
\multirow{2}{*}{\textbf{T1}} & 
\multicolumn{5}{c}{\textbf{T2}} \\
\cline{3-7}
 &  & \textbf{Speech} & \textbf{Sound} & \textbf{Singing} & \textbf{Music} & \textbf{Total} \\
\hline
Train    & 49,575  & 49,575  & 39,840 & 36,000 & 21,366 & 146,781 \\
Dev      & 49,734  & 49,734  & 19,929 & 16,000 & 5,406  & 91,069 \\
Progress & 29,269  & 28,813  & 5,729  & 4,872  & 6,461  & 45,875 \\
Eval     & 146,346 & 144,078 & 28,593 & 24,332 & 32,370 & 229,373 \\
\hline
\end{tabular}
\end{table*}
\begin{table}[t]
\centering
\caption{Real-source (T1R) and fake-source (T1F) conditions in the AT-ADD Track 1 train/dev subset. IDs continue across the Track~1 tables.}
\label{tab:t1_train_dev}
\begin{tabular}{clrr}
\hline
\textbf{ID} & \textbf{Source / Generator} & \textbf{Train} & \textbf{Dev} \\
\hline
T1R1 & Internal & 2,000 & 2,000 \\
T1R2 & AISHELL-3~\cite{shi2020aishell} & 500 & 500 \\
T1R3 & LibriTTS-R~\cite{koizumi2023libritts} & 2,500 & 2,500 \\
T1R4 & LJSpeech~\cite{ljspeech} & 2,500 & 2,500 \\
T1R5 & Common Voice~\cite{ardila2020common} & 2,499 & 2,500 \\
\hline
T1F1 & ProDiff~\cite{huang2022prodiff} & 1,999 & 1,998 \\
T1F2 & PortaSpeech~\cite{ren2021portaspeech} & 1,996 & 1,994 \\
T1F3 & DiffSpeech~\cite{liu2022diffsinger} & 1,998 & 1,999 \\
T1F4 & FastSpeech2~\cite{ren2021fastspeech} & 1,998 & 1,998 \\
T1F5 & Kokoro~\cite{nayak2025kokoro} & 2,000 & 1,998 \\
T1F6 & WaveNet~\cite{oord2016wavenet} & 1,999 & 2,000 \\
T1F7 & FastDiff~\cite{huang2022fastdiff} & 1,996 & 1,994 \\
T1F8 & MeloTTS~\cite{zhao2023melotts} & 1,996 & 1,988 \\
T1F9 & CosyVoice~\cite{du2024cosyvoice} & 1,835 & 2,000 \\
T1F10 & Parler-TTS~\cite{lacombe2024parlertts} & 1,998 & 1,995 \\
T1F11 & GradTTS~\cite{popov2021gradtts} & 1,994 & 1,997 \\
T1F12 & FastPitch~\cite{lancucki2021fastpitch} & 1,998 & 1,997 \\
T1F13 & Tacotron2~\cite{shen2018tacotron2} & 1,994 & 1,998 \\
T1F14 & Glow-TTS~\cite{kim2020glow} & 2,000 & 2,000 \\
T1F15 & WaveGlow~\cite{prenger2019waveglow} & 1,997 & 1,997 \\
T1F16 & MelGAN~\cite{kumar2019melgan} & 1,997 & 2,000 \\
T1F17 & Tortoise-TTS~\cite{betker2023scaling} & 1,992 & 1,989 \\
T1F18 & StarGANv2-VC~\cite{li2021starganv2vc} & 1,999 & 1,999 \\
T1F19 & Llasa1B~\cite{ye2025llasa} & 1,793 & 1,795 \\
T1F20 & Index-TTS~\cite{deng2025indextts} & 1,997 & 1,998 \\
\hline
-- & \textbf{Total} & \textbf{49,575} & \textbf{49,734} \\
\hline
\end{tabular}
\end{table}

\begin{table}[t]
\centering
\caption{Real-source (T1R) and fake-source (T1F) conditions in the AT-ADD Track 1 progress/eval subset.}
\label{tab:t1_eval}
\footnotesize
\setlength{\tabcolsep}{4pt}
\begin{tabular}{clrr}
\hline
\textbf{ID} & \textbf{Source / Generator} & \textbf{Progress} & \textbf{Eval} \\
\hline
T1R1 & Internal & 974 & 5,000 \\
T1R2 & AISHELL-3~\cite{shi2020aishell} & 391 & 2,000 \\
T1R3 & LibriTTS-R~\cite{koizumi2023libritts} & 832 & 3,999 \\
T1R5 & Common Voice~\cite{ardila2020common} & 1,022 & 5,000 \\
T1R6 & 3D-Speaker~\cite{zheng20233d} & 385 & 1,999 \\
T1R7 & EchoFake~\cite{zhang2025echofake} & 396 & 2,000 \\

\hline
T1F21 & ChatTTS~\cite{2noise2024chattts} & 793 & 3,967 \\
T1F22 & DiffGANTTS~\cite{liu2022diffgantts} & 1,598 & 7,992 \\
T1F23 & HiFiGAN~\cite{kong2020hifi} & 2,392 & 11,961 \\
T1F24 & BigVGAN~\cite{lee2022bigvgan} & 3,987 & 19,933 \\
T1F25 & MBMelGAN~\cite{yang2021multi} & 790 & 3,952 \\
T1F26 & ParallelWaveGAN~\cite{yamamoto2020parallel} & 798 & 3,989 \\
T1F27 & StyleMelGAN~\cite{mustafa2021stylemelgan} & 796 & 3,979 \\
T1F28 & OpenVoice2~\cite{qin2023openvoice} & 772 & 3,861 \\
T1F29 & SeedVC1~\cite{liu2024zeroshotvc} & 799 & 3,996 \\
T1F30 & StyleTTS2~\cite{li2023styletts2} & 770 & 3,852 \\
T1F31 & StyleSpeech~\cite{min2021metastylespeech} & 768 & 3,841 \\
T1F32 & VITS~\cite{kim2021conditional} & 498 & 2,489 \\
T1F33 & GPTSoVITS~\cite{gptsovits2024} & 775 & 3,875 \\
T1F34 & CosyVoice2~\cite{du2024cosyvoice2} & 784 & 3,919 \\
T1F35 & CosyVoice3~\cite{du2025cosyvoice3} & 796 & 3,981 \\
T1F36 & Fish ~\cite{liao2024fishspeech} & 696 & 3,479 \\
T1F37 & F5TTS~\cite{chen2025f5tts} & 794 & 3,971 \\
T1F38 & E2TTS~\cite{eskimez2024e2tts} & 793 & 3,965 \\
T1F39 & SparkTTS~\cite{wang2025sparktts} & 752 & 3,761 \\
T1F40 & Llasa3B~\cite{ye2025llasa} & 748 & 3,739 \\
T1F41 & Llasa8B~\cite{ye2025llasa} & 778 & 3,888 \\
T1F42 & FireRedTTS2~\cite{guo2024fireredtts} & 735 & 3,674 \\
T1F43 & IndexTTS1.5~\cite{indextts2024v15} & 797 & 3,983 \\
T1F44 & IndexTTS2~\cite{zhou2025indextts2} & 661 & 3,306 \\
T1F45 & StepAudioTTS~\cite{huang2025stepaudio} & 657 & 3,286 \\
T1F46 & EchoFake\_AIGC\_Replay~\cite{zhang2025echofake} & 398 & 1,988 \\
T1F47 & Internal\_AIGC\_Replay & 344 & 1,721 \\
\hline
-- & \textbf{Total} & \textbf{29,269} & \textbf{146,346} \\
\hline
\end{tabular}
\end{table}

\subsection{Robust Speech Deepfake Detection}

We construct the AT-ADD Track~1 dataset with predefined training, development, and evaluation splits. The evaluation split is reserved for testing and consists of real speech collected from diverse recording domains, together with fake speech generated by methods unseen in the training and development sets. It therefore evaluates CM robustness under domain shifts involving recording devices, acoustic environments, and signal perturbations. Tables~\ref{tab:t1_train_dev} and~\ref{tab:t1_eval} assign T1R IDs to real-source conditions and T1F IDs to fake-source conditions, enabling source-level analysis of both classes.

\textbf{Real speech in train/dev sets.}
The real speech subset comprises multilingual utterances from internal recordings, AISHELL-3~\cite{shi2020aishell}, LibriTTS-R~\cite{koizumi2023libritts}, LJSpeech~\cite{ljspeech}, and Common Voice~\cite{ardila2020common}. These sources provide diverse speakers, languages, recording devices, and acoustic conditions.

\textbf{Fake speech in the train/dev sets.}
The fake speech subset covers text-to-speech (TTS), voice conversion (VC), and vocoder-based resynthesis. TTS inputs are selected from the real speech data described above. VC and reference-conditioned TTS use a shared pool of real reference speakers for training and development and a disjoint pool for evaluation to prevent speaker-information leakage; reference and source speakers are also constrained to differ.

\textbf{Real speech in the Eval set.}
The real evaluation speech combines internally collected recordings captured with devices such as smart glasses, mobile phones, and in-vehicle systems with 3D-Speaker~\cite{zheng20233d}, EchoFake~\cite{zhang2025echofake}, AISHELL-3, LibriTTS-R, and Common Voice. Approximately 20\% of these utterances undergo volume scaling with factors from 0.2 to 0.9, speed perturbation with factors from 0.5 to 2.5, resampling to 8~kHz, or a combination of the three transformations. Across all splits, the real speech covers 11,299 unique speakers. Treating train/dev as a shared development pool, its speakers are disjoint from evaluation for every source except Common Voice. Although Common Voice clips were drawn from separate source partitions, the Common Voice partitioning does not enforce speaker-level separation; consequently, 18 identities overlap in the directly selected Common Voice subsets.

\textbf{Fake speech in the Eval set.}
The fake evaluation speech covers 27 \emph{unseen} TTS, VC, vocoder, and replay conditions, as listed in Table~\ref{tab:t1_eval}. Approximately 23\% undergo secondary volume scaling, speed perturbation, 8-kHz resampling, or combined transformations using the same parameter ranges as the real subset. Generator, source, generation-task, and sample-level transformation annotations are provided in the released metadata.

\begin{table}[t]
\centering
\caption{Real-source (T2R) and fake-source (T2F) conditions in the AT-ADD Track 2 train/dev subset. IDs continue across the Track~2 tables.}
\label{tab:t2_traindev}
\begin{tabular}{lclrr}
\hline
\textbf{Type} & \textbf{ID} & \textbf{Source / Generator} & \textbf{Train} & \textbf{Dev} \\
\hline

\multirow{4}{*}{Sound}
& T2R8 & AudioCaps~\cite{kim2019audiocaps} & 9,854 & 4,940 \\
& T2F46 & AudioLDM~\cite{liu2023audioldm} & 9,995 & 4,997 \\
& T2F47 & AudioLDM2~\cite{liu2024audioldm2} & 10,000 & 5,000 \\
& T2F48 & Audiogen~\cite{kreuk2022audiogen} & 9,991 & 4,992 \\
\cline{2-5}
& -- & \textbf{Total} & \textbf{39,840} & \textbf{19,929} \\

\hline

\multirow{5}{*}{Singing}
& T2R9 & OpenCpop~\cite{wang2022opencpop} & 3,414 & -- \\
& T2R10 & M4Singer~\cite{zhang2022m4singer} & 5,586 & 4,000 \\
& T2F49 & So-VITS-SVC\_inhouse~\cite{sovits2023svc} & 9,000 & 4,000 \\
& T2F50 & NeuCoSVC\_inhouse~\cite{sha2024neural} & 9,000 & 4,000 \\
& T2F51 & SeedVC~\cite{liu2024zeroshotvc} & 9,000 & 4,000 \\
\cline{2-5}
& -- & \textbf{Total} & \textbf{36,000} & \textbf{16,000} \\

\hline

\multirow{5}{*}{Music}
& T2R11 & MusicCaps~\cite{agostinelli2023musiclm} & 4,297 & 536 \\
& T2F52 & MusicGen~\cite{copet2024simple} & 4,212 & 1,204 \\
& T2F53 & MusicLDM~\cite{chen2024musicldm} & 4,276 & 1,221 \\
& T2F54 & AudioLDM2~\cite{liu2024audioldm2} & 4,289 & 1,222 \\
& T2F55 & Stable Audio open~\cite{evans2025stableaudioopen} & 4,292 & 1,223 \\
\cline{2-5}
& -- & \textbf{Total} & \textbf{21,366} & \textbf{5,406} \\

\hline
\end{tabular}
\end{table}

\begin{table}[t]
\centering
\caption{Real-source (T2R) and fake-source (T2F) conditions in the AT-ADD Track 2 progress/eval subset.}
\label{tab:t2_eval}
\begin{tabular}{lclrr}
\hline
\textbf{Type} & \textbf{ID} & \textbf{Source / Generator} & \textbf{Progress} & \textbf{Eval} \\
\hline

\multirow{9}{*}{Sound}
& T2R8 & AudioCaps~\cite{kim2019audiocaps} & 999 & 4,938 \\
& T2R12 & AVQA~\cite{yang2022avqa} & 200 & 1,000 \\
& T2R13 & CompA-R~\cite{ghosh2024gama} & 198 & 990 \\
& T2R14 & VocalSound~\cite{gong2022vocalsound} & 198 & 992 \\
& T2R15 & TUT2016~\cite{mesaros2016tut} & 200 & 1,000 \\
& T2F56 & UniAudio2~\cite{yang2026uniaudio2} & 942 & 4,711 \\
& T2F57 & Auffusion~\cite{xue2024auffusion} & 1,000 & 5,000 \\
& T2F58 & Both-Ears-Wide-Open~\cite{sun2024bothears} & 992 & 4,962 \\
& T2F59 & MeanAudio~\cite{li2025meanaudio} & 1,000 & 5,000 \\
\cline{2-5}
& -- & \textbf{Total} & \textbf{5,729} & \textbf{28,593} \\

\hline

\multirow{7}{*}{Singing}
& T2R10 & M4Singer~\cite{zhang2022m4singer} & 946 & 4,709 \\
& T2R16 & KiSing~\cite{shi2024singing} & 59 & 291 \\
& T2F60 & Free-SVC~\cite{ferreira2025freesvc} & 1,000 & 5,000 \\
& T2F61 & KNN-SVC~\cite{shao2025knnsvc} & 1,000 & 5,000 \\
& T2F62 & VEVO1.5~\cite{openmmlab2024vevosing} & 1,000 & 5,000 \\
& T2F63 & So-VITS-SVC-FSD~\cite{xie2024fsd} & 716 & 3,579 \\
& T2F64 & RVC-FSD~\cite{xie2024fsd} & 151 & 753 \\
\cline{2-5}
& -- & \textbf{Total} & \textbf{4,872} & \textbf{24,332} \\

\hline

\multirow{7}{*}{Music}
& T2R11 & MusicCaps~\cite{agostinelli2023musiclm} & 91 & 537 \\
& T2R17 & FMA~\cite{defferrard2016fma} & 1,000 & 4,978 \\
& T2R18 & FortisAVQA~\cite{ma2025fortisavqa} & 996 & 4,981 \\
& T2F65 & Musictango~\cite{melechovsky2024mustango} & 1,104 & 5,521 \\
& T2F66 & Uniaudio2~\cite{yang2026uniaudio2} & 1,085 & 5,426 \\
& T2F67 & SSM-TTM~\cite{lee2026statemusic}  & 1,104 & 5,521 \\
& T2F68 & ACE-Step~\cite{gong2025acestep} & 1,081 & 5,406 \\
\cline{2-5}
& -- & \textbf{Total} & \textbf{6,461} & \textbf{32,370} \\

\hline
\end{tabular}
\end{table}

\subsection{All-Type Audio Deepfake Detection}

In this section, we describe the composition of the proposed AT-ADD Track~2 dataset across four audio types. Table~\ref{tab:t2_traindev} summarizes the training and development sets, while Table~\ref{tab:t2_eval} presents the progress and evaluation sets.

\textbf{Speech.}
Track~2 first reserves T2F1--T2F45 for speech so that its speech-generator sequence remains aligned with Track~1. Specifically, the shared speech train/dev generators are assigned T2F1--T2F20 in the same order as T1F1--T1F20, and the 25 shared progress/eval generators are assigned T2F21--T2F45 in the same order as T1F21--T1F45. Likewise, T2R1--T2R7 denote the seven speech sources in the same order as T1R1--T1R7. Track~1's signal-perturbation and replay conditions (T1F46 and T1F47) are excluded because Track~2 emphasizes cross-type generalization rather than robustness to speech degradations. Non-speech numbering therefore begins at T2F46 and T2R8: Tables~\ref{tab:t2_traindev} and~\ref{tab:t2_eval} continue these sequences for the sound, singing, and music conditions.

\textbf{Sound.}
The sound subset is constructed from AudioCaps~\cite{kim2019audiocaps}, whose real audio is divided into non-overlapping training, development, and evaluation sets. Synthetic training and development samples are generated by text-to-audio (TTA) models conditioned on the corresponding descriptions, whereas evaluation fakes use four \emph{unseen} methods. The real evaluation set additionally includes out-of-distribution (OOD) audio from AVQA~\cite{yang2022avqa}, CompA-R~\cite{ghosh2024gama}, VocalSound~\cite{gong2022vocalsound}, and TUT2016~\cite{mesaros2016tut}.

\textbf{Singing Voice.}
The singing voice subset draws real recordings from OpenCpop~\cite{wang2022opencpop} and M4Singer~\cite{zhang2022m4singer} for training, M4Singer for development, and M4Singer and KiSing~\cite{shi2024singing} for evaluation. Fake training and development samples are generated through singing voice conversion with strictly non-overlapping source and target singers to avoid identity leakage. Evaluation fakes are produced by five \emph{unseen} methods to assess cross-model generalization.

\textbf{Music.}
The music subset is derived from MusicCaps~\cite{agostinelli2023musiclm}, with non-overlapping real training, development, and evaluation partitions. Synthetic training and development samples are generated by text-to-music (TTM) models conditioned on the corresponding descriptions, whereas evaluation fakes use four \emph{unseen} methods. The real evaluation set additionally introduces OOD music from FMA~\cite{defferrard2016fma} and FortisAVQA~\cite{ma2025fortisavqa}.
\section{Baselines}
\label{sec:baseline}

To facilitate fair comparison and lower the entry barrier, we provide a set of official baselines covering conventional CMs, SSL-based CMs, and ALLM-based CMs. These baselines span different modeling paradigms and serve as strong and reproducible starting points for participants.

\begin{table*}[t]
\centering
\caption{Baseline performance (\%) on the AT-ADD benchmarks. \textbf{Best results are highlighted in bold.}}
\label{tab:baseline_results}
\resizebox{\textwidth}{!}{
\begin{tabular}{llccc|ccc|cccc|cccc}
\toprule
\multirow{2}{*}{Type} & \multirow{2}{*}{Model} 
& \multicolumn{3}{c|}{T1} 
& \multicolumn{3}{c|}{T2} 
& \multicolumn{4}{c|}{T2 Progress (by Type)} 
& \multicolumn{4}{c}{T2 Eval (by Type)} \\
\cmidrule(r){3-5} \cmidrule(r){6-8} \cmidrule(r){9-12} \cmidrule(l){13-16}
& 
& Dev & Progress & Eval
& Dev & Progress & Eval
& Speech & Sound & Singing & Music
& Speech & Sound & Singing & Music \\
\midrule

\multirow{2}{*}{Conventional} 
& Spec-ResNet        
& 81.85 & 47.93 & 47.41 
& 57.51 & 53.22 & 53.83 
& 51.08 & 52.92 & 48.41 & 60.48 
& 51.79 & 54.35 & 49.29 & 59.88 \\

& AASIST             
& 94.16 & 60.78 & 60.39 
& 93.63 & 62.38 & 62.21 
& 63.69 & 56.88 & 64.08 & 64.87 
& 63.58 & 56.62 & 63.81 & 64.85 \\

\midrule

\multirow{2}{*}{SSL-based} 
& FT-XLSR-AASIST     
& \textbf{99.70} & \textbf{76.98} & \textbf{76.73} 
& \textbf{98.48} & \textbf{79.25} & \textbf{79.47} 
& \textbf{79.43} & \textbf{66.08} & \textbf{96.33} & \textbf{75.17} 
& \textbf{79.50} & \textbf{66.82} & \textbf{96.30} & \textbf{75.28} \\

& WPT-XLSR-AASIST    
& 96.27 & 73.56 & 73.35 
& 95.00 & 66.59 & 66.68 
& 69.42 & 52.97 & 79.81 & 64.17 
& 69.31 & 53.83 & 79.56 & 64.04 \\

\midrule

\multirow{2}{*}{ALLM-based} 
& Qwen2.5-Omni-3B    
& 93.97 & 68.65 & 68.02 
& 94.44 & 63.42 & 63.23 
& 69.47 & 50.38 & 66.31 & 67.52 
& 68.74 & 50.41 & 65.78 & 67.99 \\

& Qwen2.5-Omni-7B    
& 95.93 & 69.19 & 68.64 
& 94.70 & 61.48 & 61.78 
& 69.89 & 45.28 & 68.04 & 63.77 
& 69.29 & 45.94 & 68.03 & 63.87 \\

\bottomrule
\end{tabular}
}
\end{table*}
\subsection{Baseline Models}
We provide official implementations for all baseline systems used in AT-ADD, including conventional and SSL-based baselines\footnote{\url{https://github.com/xieyuankun/AT-ADD-Baseline}} as well as the ALLM-based baseline\footnote{\url{https://github.com/yangchunmian123/AT-ADD-ALLM-Baseline}}. We next give a brief introduction to these baseline models.

\textbf{Conventional CMs.}
These models follow the traditional pipeline of feature extraction and discriminative classification, representing standard approaches in audio deepfake detection. Although generally weaker than recent SSL-based methods, they serve as important reference systems for evaluating robustness and cross-type generalization.

\begin{itemize}[leftmargin=*]
    \item \textbf{Spec-ResNet}: A spectrogram-based detector with a ResNet backbone~\cite{he2016deep}, representing traditional feature-based CMs. In this baseline, we use only STFT-based spectrograms as input features, without incorporating specialized speech features such as Mel-spectrograms. This design aims to investigate whether simple, generic spectral representations can provide robustness to noise and support cross-type generalization.

    \item \textbf{AASIST~\cite{jung2022aasist}}: A raw waveform-based model employing a sinc convolution front-end~\cite{ravanelli2018speaker} and residual blocks, followed by spectral--temporal attention for classification. This baseline operates directly on raw waveforms without explicit feature engineering, allowing us to study the performance of end-to-end waveform-based detection.
\end{itemize}

\textbf{SSL-based CMs.}
To improve robustness and generalization, we further include models enhanced by SSL representations. These approaches leverage large-scale pre-training and have become the dominant paradigm in modern CM systems, demonstrating strong performance.

\begin{itemize}[leftmargin=*]
    \item \textbf{FT-XLSR-AASIST~\cite{tak2022automatic}}: An enhanced AASIST model using self-supervised Wav2Vec2-XLSR representations\footnote{https://huggingface.co/facebook/wav2vec2-xls-r-300m} as the front-end, providing improved transferability across languages and recording domains~\cite{phukan2024heterogeneity,pascu24_interspeech}. FT denotes full fine-tuning of all XLSR and AASIST layers. It remains a competitive baseline for audio deepfake detection.

    \item \textbf{WPT-XLSR-AASIST~\cite{xie2025detect}}: A strengthened SSL-based baseline incorporating wavelet prompt tuning (WPT) to capture frequency-invariant artifacts. By only optimizing a small number of prompt tokens, this method achieves strong performance with minimal training cost.
\end{itemize}

\textbf{ALLM-based CMs.}
We further provide ALLM-based CMs built on the Qwen audio family. These models take audio, optionally with textual instructions, as input and generate textual outputs, which are adapted into binary real/fake predictions via supervised fine-tuning (SFT). Compared with conventional CMs, ALLM-based approaches leverage large-scale multimodal pre-training and show strong potential for cross-type generalization and unified modeling across heterogeneous audio types.

\begin{itemize}[leftmargin=*]
    \item \textbf{Qwen2.5-Omni-3B / 7B}\footnote{https://huggingface.co/Qwen/Qwen2.5-Omni-3B}\footnote{https://huggingface.co/Qwen/Qwen2.5-Omni-7B}: Unified multimodal models that support audio inputs and can be adapted for audio deepfake detection via supervised fine-tuning (SFT). Compared to conventional CMs, ALLM-based CMs are capable of producing deterministic predictions and can further provide interpretable reasoning through techniques such as reinforcement learning. Recent studies have demonstrated the superior performance of ALLM-based approaches in the field of ADD~\cite{xie2026interpretable, xue2026unifying, guo2026towards}. Their potential for improving robustness and enabling unified modeling across all audio types makes them a promising direction for further exploration in this challenge.
\end{itemize}

\subsection{Baseline Performance}
Table~\ref{tab:baseline_results} reports all baselines under the official evaluation metrics. Conventional and SSL-based CMs use a fixed decision threshold of 0.5 for real/fake classification.

SSL-based CMs achieve the strongest aggregate performance on both tracks. FT-XLSR-AASIST obtains $76.73\%$ on Track~1 Eval and $79.47\%$ on Track~2 Eval, consistently outperforming the other baselines. Its Track~2 results nevertheless vary substantially by audio type: $79.50\%$ on speech, $66.82\%$ on sound, $96.30\%$ on singing, and $75.28\%$ on music. Thus, strong aggregate transfer does not imply uniform cross-type generalization.

Conventional CMs remain substantially weaker. Spec-ResNet achieves $47.41\%$ on Track~1 Eval and $53.83\%$ on Track~2 Eval, while AASIST improves these scores to $60.39\%$ and $62.21\%$, respectively, but still trails the SSL-based systems.

The general-purpose ALLM baselines are competitive on Track~1 but do not close the gap to FT-XLSR-AASIST. Qwen2.5-Omni-7B achieves $68.64\%$ on Track~1 Eval and $61.78\%$ on Track~2 Eval, while the 3B variant reaches $68.02\%$ and $63.23\%$, respectively. Their type-wise results are also uneven, particularly on sound, indicating that unified multimodal pre-training alone does not ensure balanced all-type detection.

Overall, Track~1 primarily exposes robustness limitations under unseen generators and realistic variation in real speech, whereas Track~2 reveals persistent cross-type gaps. The contrast between sound or music and singing shows that high average performance can still conceal substantial type-specific weaknesses.

\section{Evaluation Protocol and Challenge Rules}
\label{sec:evaluation}

We define the official metrics and common rules for both tracks.

\subsection{Evaluation Metrics}

The performance of submitted systems is evaluated using the $F1$-score. To ensure fair comparison under class imbalance and across audio types, Macro-$F1$ is adopted with different aggregation strategies for the two tracks.

For a given class $c$, the $F1$-score is defined as the harmonic mean of precision $P_c$ and recall $R_c$:
\begin{equation}
F1_c = \frac{2 \cdot P_c \cdot R_c}{P_c + R_c}
\end{equation}
where
\begin{equation}
P_c = \frac{TP_c}{TP_c + FP_c}, \quad
R_c = \frac{TP_c}{TP_c + FN_c}
\end{equation}
and $TP_c$, $FP_c$, and $FN_c$ denote the numbers of true positives, false positives, and false negatives for class $c$, respectively.

\textbf{Track 1.}
For Track~1 (binary classification: \emph{real} vs.\ \emph{fake}), the official metric is the Macro-$F1$ over the two classes:
\begin{equation}
\text{Macro-}F1_{\text{T1}} = \frac{1}{2}\left(F1_{\text{real}} + F1_{\text{fake}}\right)
\end{equation}
This metric assigns equal importance to both classes, making it robust to class imbalance.

\textbf{Track 2.}
For Track~2, the evaluation accounts for both class balance and audio-type balance. Specifically, for each audio type $t$, we first compute a type-wise Macro-$F1$:
\begin{equation}
\text{Macro-}F1_t = \frac{1}{2}\left(F1_{t,\text{real}} + F1_{t,\text{fake}}\right)
\end{equation}
The final score is then obtained by averaging over all audio types:
\begin{equation}
\text{Macro-}F1_{\text{T2}} = \frac{1}{4} \sum_{t=1}^{4} \text{Macro-}F1_t
\end{equation}
where $t \in \{\text{speech}, \text{sound}, \text{singing}, \text{music}\}$.

Thus, the Track~2 metric enforces a two-level balance: equal weighting across audio types and equal weighting between \emph{real} and \emph{fake} classes within each type.

\subsection{Competition Rules}

Both tracks used a closed setting. Only the released train/dev data could support training, validation, model selection, and threshold tuning; progress/evaluation data were prohibited for adaptation or pseudo-labeling. External audio and models trained for related authenticity tasks were disallowed, while traceable general-purpose pretrained models were permitted.

Signal transformations of released audio, including noise, reverberation, codecs, resampling, and RawBoost, were allowed. Public MUSAN/RIR resources could serve only as augmentation sources. Feature-, score-, and decision-level fusion were permitted with at most five compliant subsystems.

Submissions had to be fully automatic and reproducible, without manual correction or opaque services. Top teams could be asked for method, resource, model, and inference-code declarations; leakage, rule violations, or irreproducibility could result in disqualification.

\section{Challenge Results and Participating Systems}
\label{sec:challenge_results}

\begin{figure*}[t]
\centering
\includegraphics[width=\textwidth]{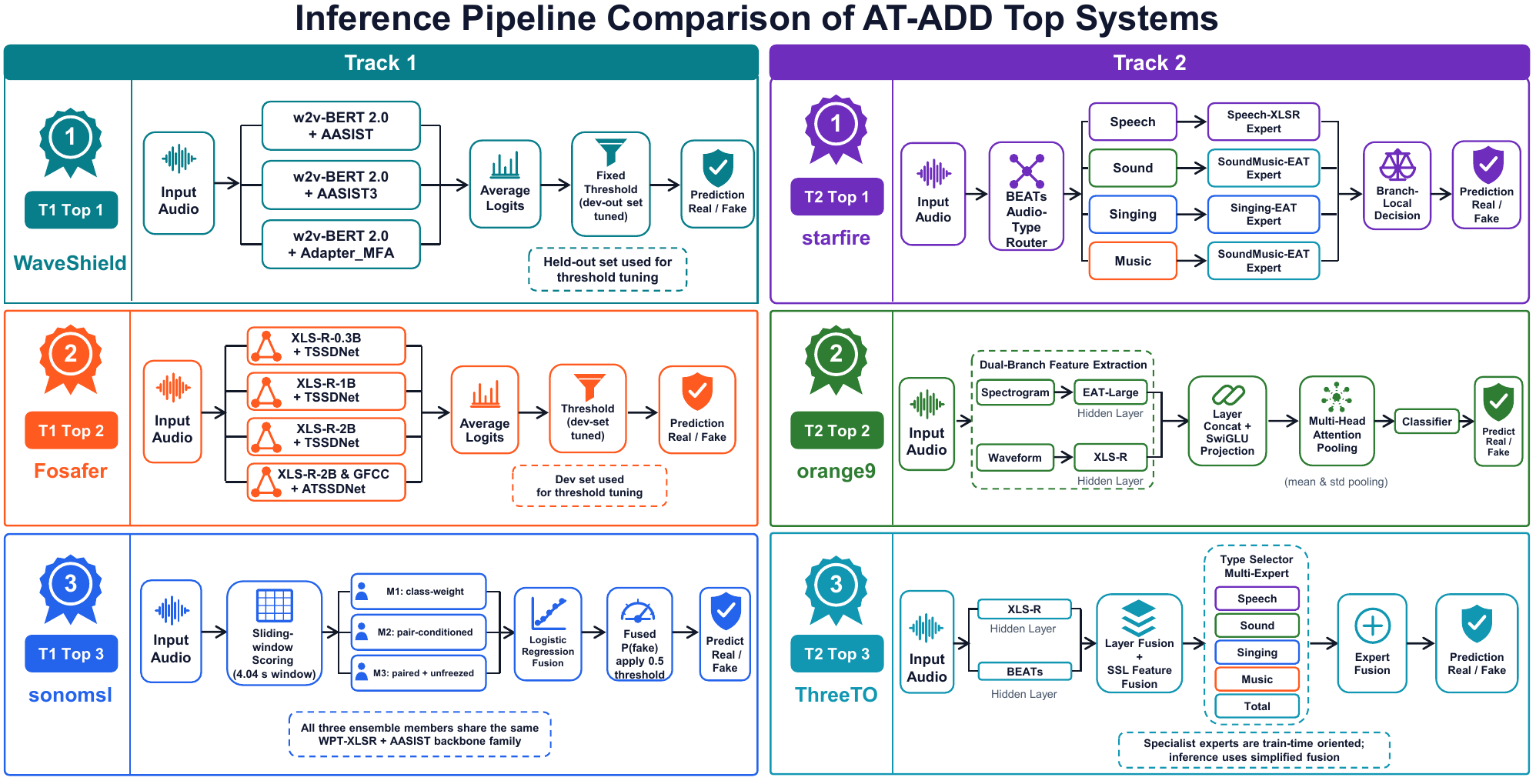}
\caption{Inference pipeline comparison of the top three systems in each AT-ADD track. Track~1 systems mainly differ in subsystem fusion, temporal scoring, and calibration, whereas Track~2 systems use either explicit audio-type routing, unified dual-representation fusion, or training-time specialists with a simplified unified inference path.}
\label{fig:systems}
\end{figure*}

After the final leaderboard freeze, we verified the submitted metadata and summarized the top five valid systems in each track in Table~\ref{tab:final_results}. We denote the system at rank $r$ in Track $t$ as T$t$-Top$r$ (e.g., T1-Top1 denotes WaveShield). The table reports the main augmentation categories, system configuration, number of submitted subsystems, progress-set score, and final evaluation score. Figure~\ref{fig:systems} complements the training-oriented information in the table by comparing the deployed inference pipelines of the top three systems in each track.

\begin{table*}[t]
\centering
\caption{Final leaderboards for Track~1 robust speech deepfake detection and Track~2 all-type audio deepfake detection. Scores are Macro-F1 (\%).}
\label{tab:final_results}
\scriptsize
\setlength{\tabcolsep}{2.5pt}
\renewcommand{\arraystretch}{0.90}
\resizebox{\textwidth}{!}{%
\begin{tabular}{llrlllrr}
\toprule
\textbf{Track} & \textbf{Rank} & \textbf{Team} & \textbf{Data Augmentation} & \textbf{System Name} & \textbf{Sub.} & \textbf{Progress} & \textbf{Eval} \\
\midrule
\multirow{5}{*}{T1} & 1 & WaveShield & Noise; Reverb; Codec; Perturb.; Replay; Seg. & w2vBERT2.0-AASIST & 3 & 84.54 & \textbf{90.71} \\
 & 2 & Fosafer & Noise; Reverb; Perturb.; Seg. & ssl\_gfcc\_multiscale\_ensemble & 4 & 85.31 & 86.67 \\
 & 3 & sonomsl & Noise; Reverb; Codec; Perturb.; Replay & sonomsl & 3 & \textbf{86.25} & 86.63 \\
 & 4 & ThreeTO & Codec; RB; Perturb.; Replay & atadd\_track1\_xlsr\_aasist\_ensemble & 3 & 83.61 & 83.79 \\
 & 5 & NKU-HLT & Reverb; Codec & EnvRobust-XLSR-AASIST & 1 & 84.20 & 83.68 \\
\midrule
\multirow{5}{*}{T2} & 1 & starfire & Codec; RB; Seg. & starfire\_track2\_audio\_type\_routed\_ensemble & 4 & \textbf{96.28} & \textbf{96.10} \\
 & 2 & orange9 & Noise; Reverb; Codec; RB; Perturb. & E2E-ATADD & 4 & 95.13 & 95.58 \\
 & 3 & ThreeTO & RB & MEDS & 3 & 93.76 & 93.95 \\
 & 4 & Fosafer & Noise; Reverb; Perturb.; Seg. & ssl\_gfcc\_scene\_adaptive\_ensemble & 5 & 91.61 & 91.62 \\
 & 5 & KETI-KU & Codec & Wav2Vec2-BERT-AASIST & 1 & 91.31 & 91.10 \\
\bottomrule
\end{tabular}}
\begin{minipage}{\textwidth}
\vspace{1mm}
\footnotesize\emph{Note:} Noise denotes augmentation using MUSAN or other public noise corpora. Reverb denotes reverberation-related augmentation, e.g., RIR. Codec denotes codec or re-encoding augmentation, e.g., g711alaw, mp3, opus, etc. Perturb. denotes signal-level perturbations, e.g., quantization, dynamic-range compression, clipping, EQ, masking, resampling, time stretching, pitch shifting, etc. Replay denotes playback simulation through loudspeaker playback. Seg. denotes segment-level operations, e.g., random crop and bonafide segment concatenation. RB denotes RawBoost.
\end{minipage}
\end{table*}

\begin{figure*}[!t]
	\centering
	\includegraphics[width=\textwidth]
	{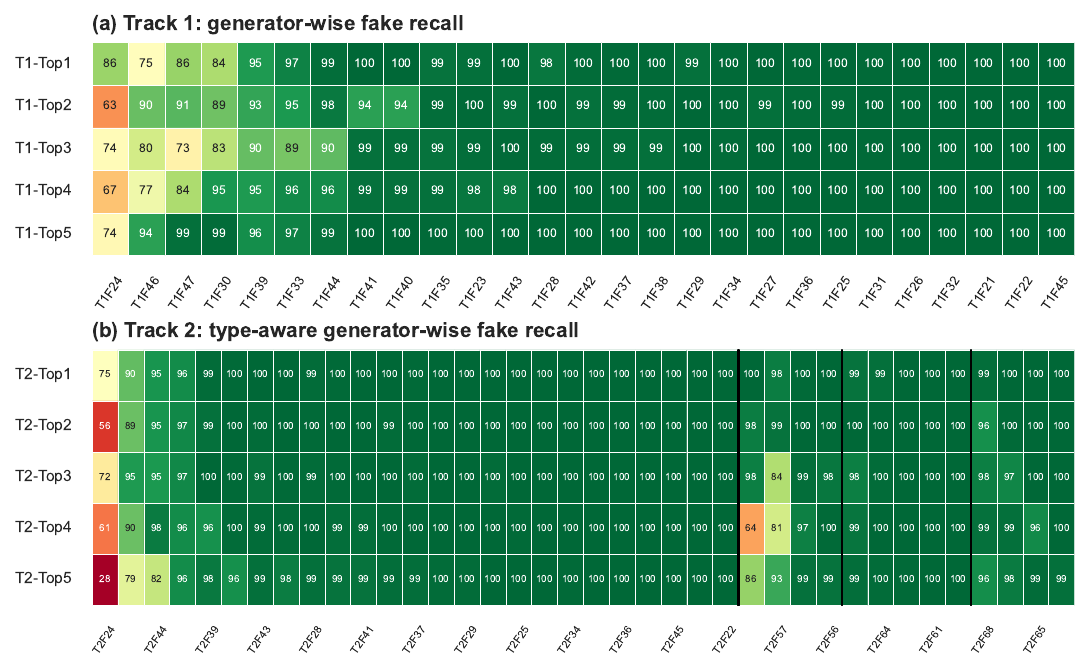}
	\caption{Generator-wise fake recall (\%) of the top five submissions.
		Cell colors span 50--100\%, and values are rounded to integers for
		readability. Columns are sorted from difficult to easy within Track~1
		and within each Track~2 audio type. Track~2 boundaries separate speech
		(T2F21--T2F45), sound (T2F56--T2F59), singing (T2F60--T2F64), and
		music (T2F65--T2F68). Generator names are given in
		Tables~\ref{tab:t1_eval} and~\ref{tab:t2_eval}.}
	\label{fig:generator_difficulty}
\end{figure*}

\subsection{Speech Deepfake Detectors}

The top Track~1 systems all relied on strong speech self-supervised learning (SSL) representations, but differed in how they controlled channel and generator mismatch and converted segment-level evidence into file-level decisions.

\textbf{T1-Top1 (WaveShield).} T1-Top1 ensembled three W2V-BERT~2.0\footnote{\url{https://huggingface.co/facebook/w2v-bert-2.0}} systems: two AASIST-based variants~\cite{jung2022aasist,borodin2024aasist3} and an Adapter-MFA branch pooling multiple encoder layers~\cite{mao2026waveshield}. \emph{Training:} it first optimized LoRA/adapters~\cite{hu2022lora} and back-ends with the pretrained encoder frozen, then performed joint and large-margin fine-tuning with focalized AM-Softmax. Diversity-balanced sampling covered generators, codecs, perturbations, replay, and bonafide concatenation; an internal held-out split determined checkpoints and thresholds. \emph{Inference:} the three systems scored the same input, their real/fake logits were averaged, and the fake--real difference was compared with one fixed held-out threshold, yielding 90.71\% Macro-F1.

\textbf{T1-Top2 (Fosafer).} T1-Top2 combined four XLSR subsystems at three model scales, using TSSDNet back-ends and one GFCC-enhanced ATSSDNet branch~\cite{babu2021xls,hua2021tssdnet,fosafer2026metadata}. \emph{Training:} every front-end and back-end was fully fine-tuned on the official training set with Binary Focal Loss. Roughly 30\% of training instances received online noise, reverberation, filtering, level, silence, or speed/pitch augmentation, and the development set was used for model selection. \emph{Inference:} the four subsystems independently generated logits, which were averaged before applying a development-selected threshold, producing 86.67\% Macro-F1.

\textbf{T1-Top3 (sonomsl).} T1-Top3 used three WPT-XLSR1B-AASIST members with a wavelet-prompt-tuned XLSR front-end~\cite{xie2025detect,vo2026shortcut}. \emph{Training:} channel, codec, replay, and device simulations were combined with weighted classification losses, generator-adversarial learning, and GroupDRO~\cite{ganin2015unsupervised,sagawa2020groupdro}. The members introduced diversity through independent versus condition-matched augmentation and different degrees of XLSR fine-tuning. \emph{Inference:} each file was divided into overlapping 4.04-second windows; window logits were median-pooled per model and the three file scores were fused by held-out regularized logistic regression. A 0.5 fused-probability threshold gave 86.63\% Macro-F1.

\textbf{T1-Top4 (ThreeTO).} T1-Top4 submitted a three-subsystem wav2vec2-XLSR-AASIST ensemble. One branch fused XLSR layers 3, 11, and 24 and replaced the original AASIST input projection with GRKAN~\cite{yang2024kat}; two CQCC-enhanced branches used SSL-to-CQCC cross-attention~\cite{todisco2016new}, with the third receiving stronger codec, RawBoost~\cite{tak2022rawboost}, pitch, resampling, speed, and replay augmentation. Inference used weighted logit fusion across the three branches.

\textbf{T1-Top5 (NKU-HLT).} T1-Top5 used a compact single EnvRobust-XLSR-AASIST model. It first fully fine-tuned wav2vec2-XLSR-300m and AASIST under codec/resampling augmentation, then froze the SSL front-end and trained with paired clean/RIR views. A score-consistency mean squared error term aligned the RIR-view and clean-view scores, and a development-set operating point produced the final prediction.

\subsection{All-Type Audio Deepfake Detectors}

Track~2 required a binary decision when the input type could be speech, sound, singing, or music. The leading systems addressed this hidden-type condition through explicit routing, unified representation fusion, or training-time specialist supervision.

\textbf{T2-Top1 (starfire).} T2-Top1 combined a BEATs audio-type router~\cite{chen2023beats,starfire2026routing} with three specialized detectors: XLSR-AASIST for speech and EAT-AASIST~\cite{chen2024eat} branches for sound/music and singing. \emph{Training:} the router learned the four official audio types with class-weighted cross-entropy. The binary experts were trained separately on type-balanced mixtures: speech used codec, replay, RawBoost, and TTS augmentation, whereas the EAT experts used longer crops and codec-augmented non-speech mixtures. Checkpoints and operating points were selected on development data. \emph{Inference:} the router assigned one type, the corresponding expert aggregated temporal crops, and a branch-local rule produced the final label, achieving 96.10\% Macro-F1.

\textbf{T2-Top2 (orange9).} T2-Top2 used two unified EAT-XLSR detectors with learned layer fusion, SwiGLU projection~\cite{shazeer2020glu}, and attentive statistical pooling~\cite{fan2026beyond}. One detector also predicted audio type, while Whisper~\cite{radford2023whisper} and an XLSR speech refiner handled difficult speech inputs. \emph{Training:} the official train/dev pool was stratified by type, label, and generator into folds. The unified models used random crops and RawBoost; the speech refiner additionally used codec, MUSAN, and RIR augmentation. \emph{Inference:} each model averaged five-crop logits, and the two unified scores were fused for every clip. The speech refiner was added only when the type head predicted speech and Whisper returned a transcription. A fixed 0.5 threshold yielded 95.58\% Macro-F1.

\textbf{T2-Top3 (ThreeTO).} Multi-Expert Audio Deepfake detection with dual Self-supervised learning frontends (MEDS) fused selected XLSR and BEATs layers before unified and type-specialist AASIST branches~\cite{liu2026meds}; a second subsystem added SSL-to-CQCC cross-attention. \emph{Training:} the specialist selected by the ground-truth audio type provided auxiliary supervision, while the unified branch learned from all samples with class-weighted cross-entropy. Both SSL encoders were fine-tuned and RawBoost was applied selectively by audio type. \emph{Inference:} only unified branch scores were retained. Multi-crop predictions were pooled using a rule and threshold chosen by a high-confidence type classifier, and the MEDS and MEDS-CQCC outputs were ensembled, producing 93.95\% Macro-F1.

\textbf{T2-Top4 (Fosafer).} T2-Top4 combined four XLSR/TSSDNet or GFCC/ATT-TSSDNet real/fake detectors~\cite{hua2021tssdnet} with an EAT-large four-class scene classifier. The detectors used online noise, reverberation, level, filtering, silence, and speed/pitch augmentation. At inference, detector scores were fused and interpreted with the scene-specific threshold selected by the auxiliary classifier.

\textbf{T2-Top5 (KETI-KU).} T2-Top5 used one fully fine-tuned W2V-BERT~2.0+AASIST detector with weighted cross-entropy. The single-system score was converted to the final binary label using a threshold selected on the development set.

\section{Submission-Level Performance Analysis}
\label{sec:submission_analysis}

\begin{table}[!t]
	\centering
	\caption{Most difficult evaluation generators, ranked by mean fake
		recall across the top five systems. The range gives the minimum and
		maximum system recall.}
	\label{tab:hard_generators}
	
	\scriptsize
	\setlength{\tabcolsep}{3pt}
	
	\begin{tabular}{cllc}
		\toprule
		\textbf{Track} & \textbf{ID} & \textbf{Generator}
		& \textbf{Mean (range), \%} \\
		\midrule
		T1 & T1F24 & BigVGAN & 72.84 (62.81--85.93) \\
		T1 & T1F46 & EchoFake replay & 83.18 (74.65--93.91) \\
		T1 & T1F47 & Internal replay & 86.69 (73.04--99.13) \\
		\midrule
		T2 & T2F24 & BigVGAN (speech) & 58.33 (28.35--75.09) \\
		T2 & T2F30 & StyleTTS~2 (speech) & 88.49 (78.64--94.69) \\
		T2 & T2F59 & MeanAudio (sound) & 89.16 (64.12--100.00) \\
		\bottomrule
	\end{tabular}
\end{table}

This section moves beyond aggregate leaderboard scores and analyzes the sample-level decisions of the top five submissions in each track. We first verified that every prediction file contained exactly one valid decision for every evaluation utterance. Re-evaluation with the released scoring implementation reproduced all ten leaderboard scores exactly. Because participants submitted hard real/fake decisions rather than continuous scores, the analysis focuses on class-wise recall, prediction disagreement, error overlap, hard-decision voting, and ranking stability; score calibration, threshold sweeps, and ROC-based metrics are not inferred from unavailable information.

For a fake generator $g$, we define generator recall as the proportion of its samples classified as fake. This is preferable to a generator-specific F1 score because a single fake generator contains no real examples. Track~2 type-wise Macro-F1 is computed using both real and fake samples within each audio type, exactly following the official protocol. Generator IDs T1F1--T1F47 and T2F1--T2F68 refer to the conditions defined in Tables~\ref{tab:t1_train_dev}--\ref{tab:t2_eval}; only evaluation-set IDs appear in the following figures.

\begin{figure*}[!t]
\centering
\includegraphics[width=\textwidth]{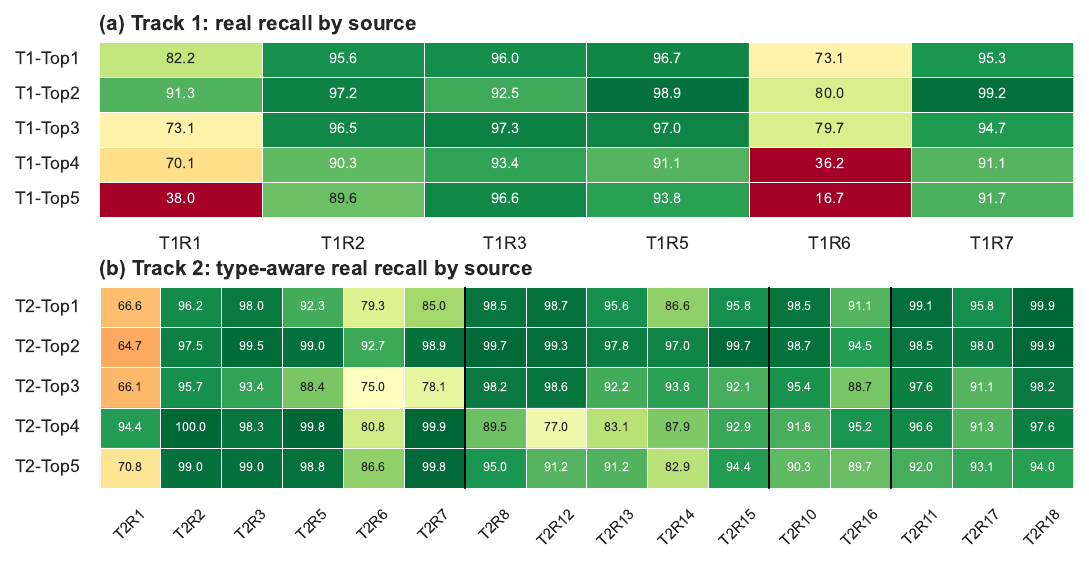}
\caption{Source-wise recall (\%) on real evaluation samples for (a) Track~1 and (b) Track~2. Cell colors span 50--100\%. Only sources present in the evaluation sets are shown; thus, T1R4 is omitted from panel (a). Columns use the real-source IDs defined in Tables~\ref{tab:t1_train_dev}--\ref{tab:t2_eval}; Track~2 boundaries separate speech, sound, singing, and music. Higher values indicate better bonafide preservation.}
\label{fig:real_recall}
\end{figure*}

\begin{figure*}[!t]
\centering
\includegraphics[width=\textwidth]{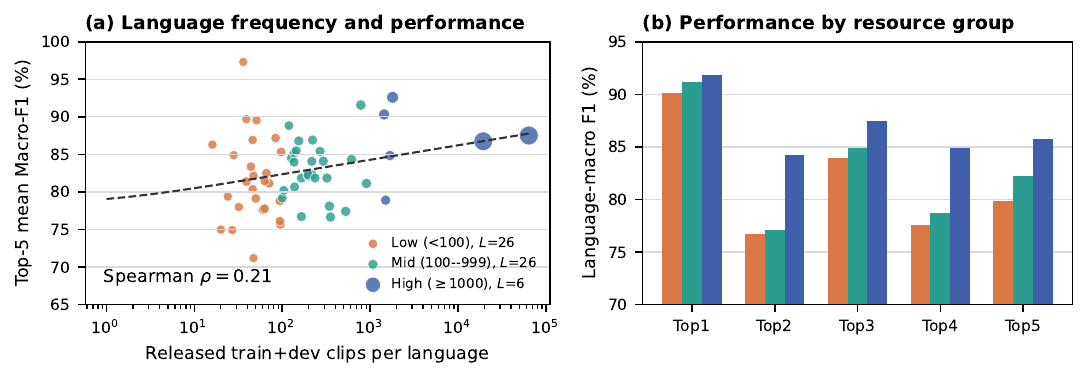}
\caption{Track~1 language-level robustness of the top five submissions. (a) Released train+dev support versus the mean language-level Macro-F1 across the five systems; marker size denotes evaluation support and the dashed line is a log-linear trend. (b) Macro-F1 averaged equally over languages within each resource group. Only languages with at least 20 real and 20 fake evaluation samples are included.}
\label{fig:language_resource}
\end{figure*}

\subsection{Generator-Level Difficulty}

Figure~\ref{fig:generator_difficulty} reveals that aggregate scores conceal a small number of concentrated failure modes. In Track~1, BigVGAN (T1F24) was the only generator for which all five systems achieved below 90\% fake recall. Its mean recall was only 72.84\%, with individual values ranging from 62.81\% to 85.93\%. The two replay conditions were the next most difficult: EchoFake replay (T1F46) averaged 83.18\%, and the internal replay condition (T1F47) averaged 86.69\%. Their wide cross-system ranges in Table~\ref{tab:hard_generators} show that replay robustness was not shared uniformly, even though replay simulation appeared in several teams' augmentation pipelines. StyleTTS~2 (T1F30) formed a second, less severe synthesis-related weakness, whereas most remaining generators were detected nearly perfectly.
Track~2 exposed a sharper version of the same generator shift. Speech BigVGAN (T2F24) averaged only 58.33\% recall and ranged from 28.35\% for T2-Top5 to 75.09\% for T2-Top1. This condition alone was markedly harder than any non-speech generator. StyleTTS~2 (T2F30) was the next hardest speech generator, while MeanAudio (T2F59) exposed a system-specific sound weakness: T2-Top1 detected all of its samples, whereas T2-Top4 detected only 64.12\%. These results indicate that ``all-type'' difficulty is not explained solely by the number of audio types; it is driven by the interaction between particular generators and the representation or routing strategy used by each detector.

\subsection{Real-Sample Detection}

The generator heatmap must be interpreted jointly with source-level real recall. Figure~\ref{fig:real_recall}(a) shows that the main Track~1 weaknesses were concentrated in the internally collected recordings (T1R1) and 3D-Speaker (T1R6), rather than being uniform across real speech. On T1R6, recall ranged from 16.71\% for T1-Top5 to 79.99\% for T1-Top2; in contrast, all systems performed substantially better on AISHELL-3 (T1R2), LibriTTS-R (T1R3), Common Voice (T1R5), and EchoFake (T1R7). T1-Top5's strong fake-generator coverage therefore came at the cost of severe rejection of challenging real recording conditions, whereas the leading systems achieved a better balance between spoof coverage and bonafide preservation.

As shown in Fig.~\ref{fig:real_recall}(b), Track~2 exhibited both type- and source-dependent real-sample recall. The widest speech variation occurred for internal recordings (T2R1), 3D-Speaker (T2R6), and EchoFake (T2R7), while the public speech sources T2R2, T2R3, and T2R5 were generally easier. For sound, the OOD AVQA (T2R12), CompA-R (T2R13), and VocalSound (T2R14) conditions separated the systems more clearly than TUT2016 (T2R15). Source effects were also visible within singing and music, although their recalls were more stable overall. These contrasts demonstrate that type-level averages can conceal substantial dataset-specific real/fake biases.

\subsection{Language-Level Robustness}

Track~1 represents 96 base languages across the released splits, of which 94 occur in evaluation. We normalize regional tags to their base language and retain the 58 evaluation languages containing at least 20 real and 20 fake samples, avoiding unstable class-balanced scores for very small subsets. Resource level is defined by the number of released train+dev clips: low ($<100$), mid (100--999), and high ($\geq1000$), containing 26, 26, and 6 analyzed languages, respectively.

As shown in Fig.~\ref{fig:language_resource}(a), language support has only a weak positive association with performance (Spearman $\rho=0.21$), with substantial variation among languages of comparable size. Nevertheless, the Top-5 mean language-macro F1 rises from 81.66\% in the low-resource group to 82.83\% and 86.83\% in the mid- and high-resource groups. Figure~\ref{fig:language_resource}(b) further shows that T1-Top1 changes only from 90.15\% to 91.83\% between the low- and high-resource groups, whereas the other systems gain 3.49--7.50 points. Thus, training-language coverage matters, but generator, source, and recording-condition composition remain strong confounders; the winning system is also notably less sensitive to language scarcity.

\subsection{Type Balance and Error Complementarity}

\begin{figure*}[!t]
	\centering
	\includegraphics[width=\textwidth]{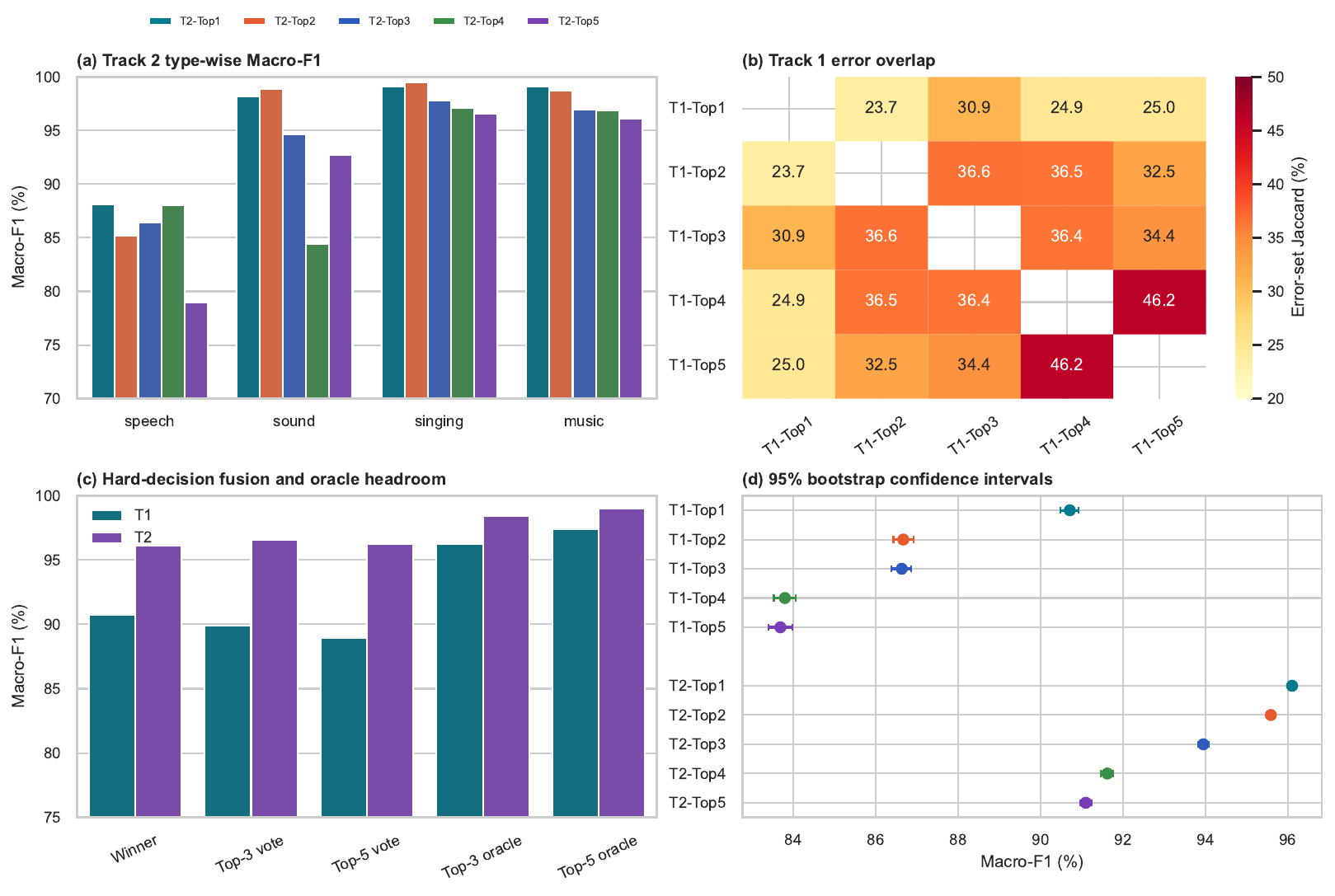}
	\caption{Submission-level diagnostics. (a) Track~2 type-wise Macro-F1. (b) Pairwise Jaccard similarity of Track~1 error sets; the diagonal is omitted. (c) Official winner, hard-decision majority voting, and non-deployable oracle upper bounds. (d) Stratified-bootstrap 95\% confidence intervals from 5,000 resamples.}
	\label{fig:submission_diagnostics}
\end{figure*}

Figure~\ref{fig:submission_diagnostics}(a) confirms that speech remained the limiting Track~2 audio type: speech Macro-F1 ranged from 78.96\% to 88.07\%, while the strongest systems exceeded 98\% on sound, singing, and music. T2-Top4 nearly matched T2-Top1 on speech but dropped sharply on sound, whereas T2-Top2 was strongest on sound and singing but trailed T2-Top1 on speech. Explicit routing and unified fusion therefore produced complementary type profiles.

To quantify complementarity, we computed the Jaccard similarity between each pair of error sets. As shown in Fig.~\ref{fig:submission_diagnostics}(b), Track~1 pairwise overlap ranged from 23.73\% to 46.22\%, demonstrating that even similarly ranked systems failed on substantially different samples. However, diversity alone did not guarantee a better deployable ensemble. Majority voting among the Track~1 top three and top five reached 89.89\% and 88.94\%, respectively, both below T1-Top1's 90.71\%. The stronger winner was diluted by correlated decisions from weaker systems. In contrast, as shown in Fig.~\ref{fig:submission_diagnostics}(c), Track~2 top-three voting improved Macro-F1 from 96.10\% to 96.55\%, while top-five voting reached 96.24\%.

We further report an oracle upper bound that marks a sample correct whenever at least one member predicts it correctly. This is not a deployable fusion rule because it uses the ground-truth label, but it measures recoverable complementarity. The Top-3 and Top-5 oracle scores were 96.25\% and 97.39\% for Track~1, and 98.40\% and 98.98\% for Track~2. The gaps between majority voting and the oracle bounds suggest that future work should learn generator- or type-conditional fusion rather than applying an unweighted vote.
\subsection{Statistical Stability of the Leaderboard}

We assessed sampling uncertainty with 5,000 nonparametric bootstrap resamples. Track~2 was stratified by audio type so that each replicate retained the official four-type aggregation. T1-Top1's 95\% interval was 90.49--90.91\%, and it ranked first in every Track~1 replicate. However, the 0.04-point gap between T1-Top2 and T1-Top3 was not stable: T1-Top2 ranked second in 62.62\% of replicates and T1-Top3 in 37.38\%. The fourth/fifth ordering was similarly uncertain, with T1-Top4 ranked fourth in 80.62\% of replicates. These results distinguish a decisive Track~1 winner from two statistically close rank pairs.

Track~2 showed a different pattern. The intervals were 96.01--96.19\%, 95.50--95.67\%, 93.82--94.07\%, 91.46--91.76\%, and 90.96--91.23\% for T2-Top1 through T2-Top5, respectively. As shown in Fig.~\ref{fig:submission_diagnostics}(d), their order was unchanged in all 5,000 replicates, indicating that the Track~2 leaderboard gaps reflect stable evaluation differences rather than sampling noise.

Together, these analyses lead to four conclusions. First, the benchmark difficulty is concentrated in a few generation conditions, especially BigVGAN and replayed speech. Second, type-aware routing and unified models exhibit complementary strengths, making conditional fusion promising for all-type detection. Third, language support explains only part of Track~1 difficulty, while the winning system is comparatively stable across resource levels. Fourth, aggregate ranks should be accompanied by uncertainty estimates: the Track~1 winner is robust, but several adjacent ranks are practically indistinguishable under resampling.

\section{Discussion}
\label{sec:discussion}
The official baselines and final submissions reveal a substantial gap between generic benchmark models and systems explicitly designed for hidden distribution shifts. On Track~1, the strongest baseline, FT-XLSR-AASIST, reached 76.73\% Macro-F1, whereas T1-Top1 achieved 90.71\%. The winning systems retained strong speech SSL front-ends but added broader channel simulation, complementary back-ends, window-level inference, and calibrated fusion. The result suggests that robustness cannot be obtained from representation scale alone: the training distribution, augmentation policy, and decision rule jointly determine transfer to unseen generators and realistic capture conditions.

On Track~2, FT-XLSR-AASIST achieved 79.47\% Macro-F1, while T2-Top1 reached 96.10\%. The largest gains came from modeling audio heterogeneity explicitly. Routed and multi-expert systems assigned different representations, pooling strategies, or thresholds to speech, sound, singing, and music, whereas unified systems benefited from combining speech-oriented and general-audio encoders. The best Track~2 result was still lowest on speech, indicating that realistic speech generation and recording-domain variability remain harder than the evaluated non-speech types even within an all-type setting.

Several common patterns emerge across both tracks. First, large-scale SSL encoders provide transferable foundations, but no single encoder dominates every audio type. Second, augmentation is most effective when organized around deployment conditions such as noise, reverberation, codec distortion, replay, and signal perturbation. Third, multi-crop aggregation and threshold calibration materially affect the final binary decision. Fourth, the sample-level analysis shows that aggregate rank alone is insufficient: BigVGAN and replayed speech concentrate errors, adjacent Track~1 ranks overlap statistically, and simple voting helps Track~2 but not Track~1. Finally, ensembles improve accuracy but increase inference cost and complicate reproducibility; compact systems such as T1-Top5 and T2-Top5 therefore remain useful reference points for practical deployment. Future evaluations should add acoustic-condition annotations, model size, latency, continuous scores, and calibration metrics to the generator- and type-level breakdowns introduced here.

\section{Conclusion}
This paper presented AT-ADD, a benchmark and challenge for robust speech and all-type audio deepfake detection. We detailed the two tasks, dataset construction, evaluation protocol, official baselines, final leaderboards, and the designs of the top-ranked systems. Sample-level analysis further localized failures by generator and audio type, quantified cross-system complementarity, and assessed the statistical stability of the rankings. The results demonstrate a clear advance beyond the baselines while exposing unresolved challenges involving unseen generators, realistic domain shifts in speech, and heterogeneous audio types. AT-ADD provides a reproducible benchmark for developing audio forensic systems that are both accurate and practically robust.


\bibliographystyle{IEEEtran}
\renewcommand{\IEEEbibitemsep}{-0.3pt}
\bibliography{myrefs}

%
%
%
%
%
%
%
%

\end{document}